\documentclass{elsart1p}

  \usepackage{graphicx}

\usepackage{amssymb}

\begin{document}

\begin{frontmatter}



\title{Spin Physics Program at RHIC-PHENIX}


\author{K. Aoki for the PHENIX Collaboration}

\address{Department of Physics, Kyoto University, Kyoto 606-8502, Japan.}
\address{RIKEN Nishina Center, RIKEN, Wako, Saitama 351-0198, Japan.}
\begin{abstract}
Longitudinal spin physics program at RHIC-PHENIX is introduced.
Recent results of $\pi^0$ cross section and $A_{LL}$ are presented and discussed.
\end{abstract}

\end{frontmatter}

\section{Introduction}
\label{introduction}
PHENIX\cite{PHENIXoverview} longitudinal spin program\cite{PHENIXSpin}
has been launched to reveal the spin structure of proton. There exists wide efforts
towards the understainding of the proton spin, mainly from polarized deep inelastic scattering (pDIS) so far.
Many polarized parton distribution functions (pol. PDF)
are suggested (e.g.\cite{Jager:2002xm}\cite{AAC}) based
on pDIS data. But there has remained large uncertainty especially on $\Delta G$,
the gluon spin contribution to the proton.

Since 2002,
longitudinally polarized protons have been collided at RHIC,
the world's first polarized $pp$ collider. 
The experiment gives us unique opportunity to explorer $\Delta G$ directory
because gluon interacts in leading order.

PHENIX has excellent gamma identification, trigger and high-rate capability,
which makes $\pi^0$ measurement feasible.
Double helicity asymmetry ($A_{LL}$) of $\pi^0$ production in polarized $pp$ collisions
is sensitive to $\Delta G$ because gluon-gluon and quark-gluon scattering dominates
in the measured $p_T$ range.
$A_{LL}$ is defined as
\begin{equation}
A_{LL} = \frac{\sigma_{++}-\sigma_{+-}}{\sigma_{++}+\sigma_{+-}}
\end{equation}
where $\sigma_{++(+-)}$ is the production cross-section in like (unlike)
helicity collisions. Theory calculation based on factorized pertubative QCD (pQCD)
with various pol. PDF models are available.
Thus we can derive information on pol.
PDF from $A_{LL}$.
Since the argument is based on pQCD, we need to confirm pQCD applicability in the measured energy.
Thus our strategy is the following: We measure cross-section for pQCD confirmation,
and then measure $A_{LL}$ to extract $\Delta G$.

Our figure of merit for $A_{LL}$ has grown
dramatically due to the improvements
in both luminosity and polarization(Table \ref{table:runhistory}).
In 2006, in addition to $\sqrt{s}=200$GeV, we took data at $\sqrt{s}=62.4$GeV
which can probe higher Bjorken-$x$ at fixed $p_T$.

\begin{center}
\begin{table}[htb]
\begin{tabular}{rrrrr} \hline
\ \ Year & \ \ \ $\sqrt{s}$ (GeV) & Figure Of Merit (nb${}^{-1}$)&
\ \ Integrated Luminosity (pb${}^{-1}$) & \ \ Polarization(\%) \\ \hline \hline
2003 & 200\ \ \ &  2.6     & 0.22\ \    & 35, 30 \\
2004 & 200\ \ \ &  2.9     & 0.075      & 45, 44 \\
2005 & 200\ \ \ & 170\ \ \ & 3.4\ \ \ \ & 47    \\
2006 & 200\ \ \ & 970\ \ \ & 7.5\ \ \ \ & 60    \\
     &  62.4   &  2.2     & (for preliminary results) 0.042  & 48  \\ \hline 
\end{tabular}
\caption{RHIC-PHENIX longitudinally polarized $pp$ collision run history.
Figure of merit (FOM) is defined as $P^4L$ where $P$ denotes polarization
and $L$ integrated luminosity. Statistical uncertainty is proportional
to $1./\sqrt{\mathrm{FOM}}$.
}
\label{table:runhistory}
\end{table}
\end{center}
\vspace{-0.8cm}
\section{Results and Discussions}
\begin{figure}
\begin{minipage}{0.4\textwidth}
\includegraphics[width=\textwidth]{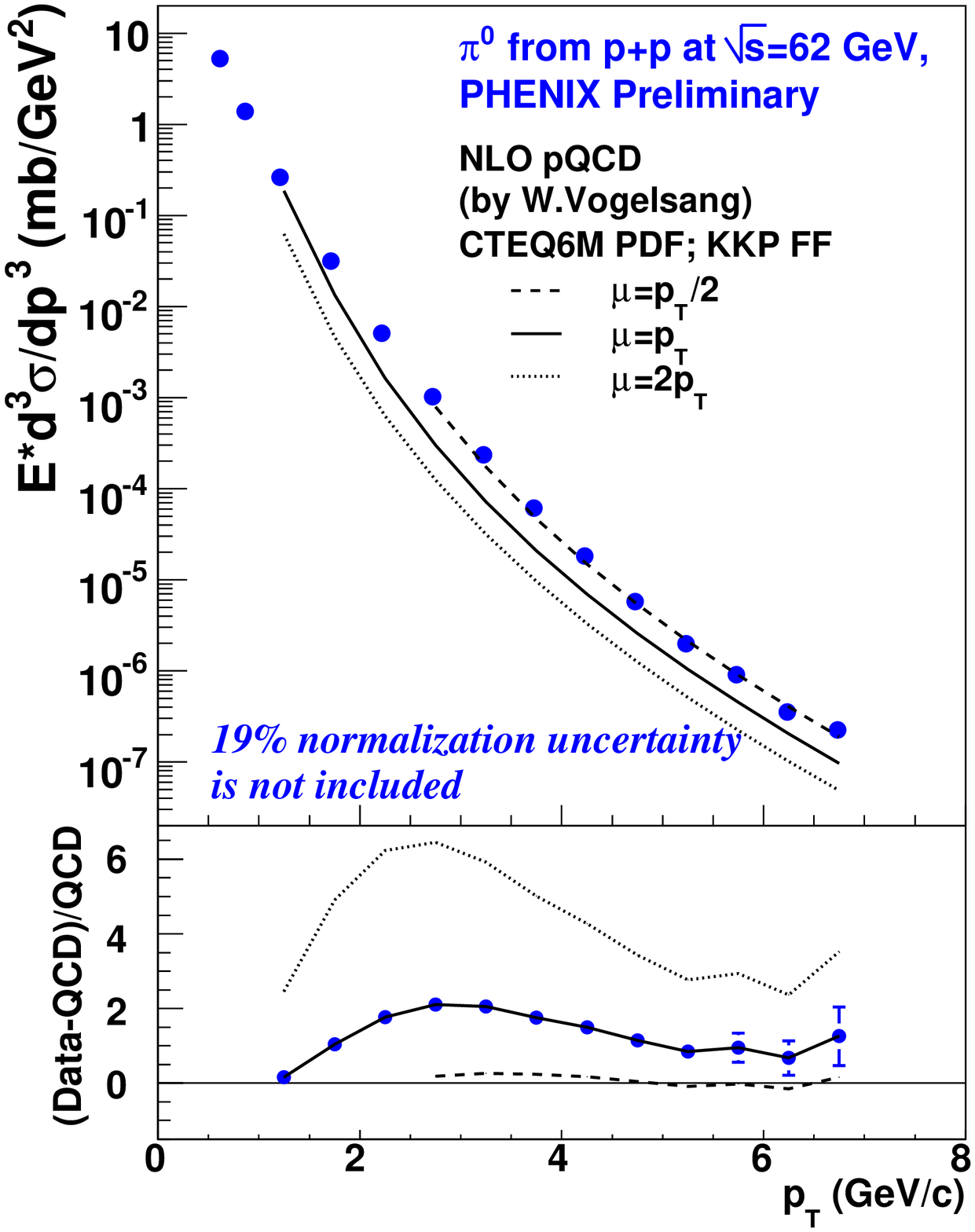}
\caption{Preliminary results of $\pi^0$ production cross section in $pp$ collisions
at $\sqrt{s}=$ 62.4GeV}
\label{fig:cs}
\end{minipage}
\begin{minipage}{0.1\textwidth}
\end{minipage}
\begin{minipage}{0.5\textwidth}
\includegraphics[width=\textwidth]{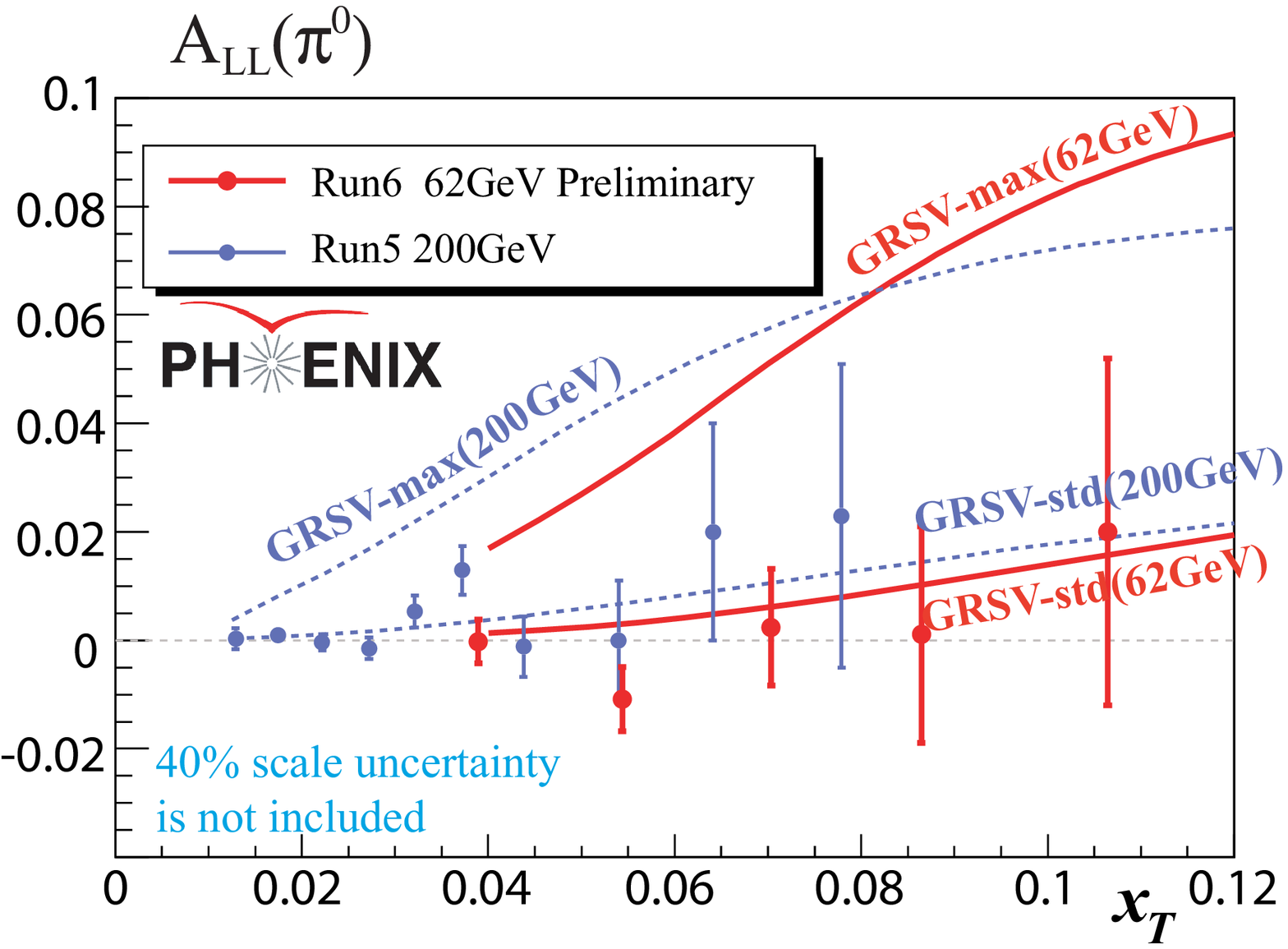}
\caption{PHENIX $A_{LL}$ $\pi^0$ results. Red: Prelminary results at $\sqrt{s}=62.4$GeV
in run 2006. Blue: Final results at $\sqrt{s}=200$GeV in 2005.}
\label{fig:allpi0_run6run5}
\end{minipage}
\end{figure}

We have published results of mid-rapidity $\pi^0$ and direct photon cross section
at $\sqrt{s}=200$GeV\cite{PHENIXpi0}\cite{PHENIXdirectphoton}.
In addition,
we present preliminary results for $\pi^0$ at $\sqrt{s}=62.4$GeV in fig. \ref{fig:cs}.
The results agree well with pQCD calculations at both energy within theoretical
uncertainties.
Thus we can discuss our $A_{LL}$ results based on pQCD.

\begin{figure}
\begin{minipage}{0.4\textwidth}
\includegraphics[width=\textwidth]{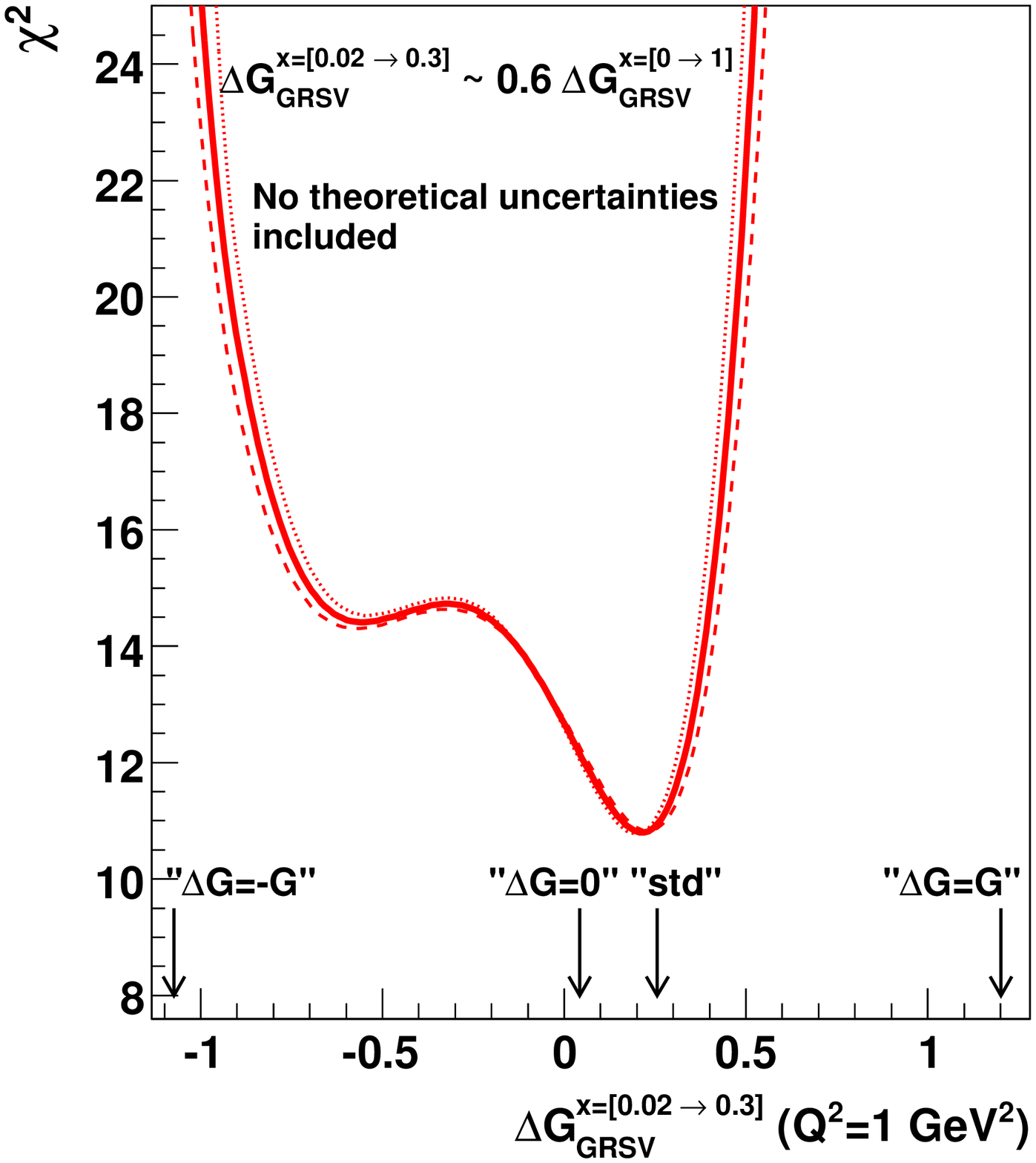}
\end{minipage}
\begin{minipage}{0.6\textwidth}
\caption{
The $\chi^2$ distribution of the measured $A_{LL}$ results at $\sqrt{s}=$200GeV
in run 2005 versus the value of the first moment
of the polarized gluon distribution (solid line)
in the $x$ range from 0.02 to 0.3 for which the $\pi^0$ results are sensitive.
Dotted and dashed curves correspond to polarization scale uncertainty of $\pm 9.4$\%.
Arrows indecate four different pol. PDF models. Our results
reject two large polarization senarios ($\Delta G=\pm G$) and prefer small $\Delta G$.
The two minima indecate the contribution from gluon-gluon scattering
which hides the sign of $\Delta G$.
}
\label{fig:chi2}
\end{minipage}
\end{figure}

The blue points in fig. \ref{fig:allpi0_run6run5} shows $\pi^0$ $A_{LL}$ published results
in run 2005 at $\sqrt{s}=$200GeV\cite{PHENIXpi0} as a function of $x_T=2p_T/\sqrt{s}$.
It is overlayed with pQCD calculation using two pol. PDF models.\cite{Jager:2002xm}
More sophisticated comparison\cite{PHENIXpi0} with
theroy was performed as in fig. \ref{fig:chi2}.
The results reject large gluon polarization senarios ($\Delta G=\pm G$)
and prefer small $\Delta G$. Another independent analysis including
our data likewise indicates $\Delta G$ is not large\cite{AAC}.

The red points in fig. \ref{fig:allpi0_run6run5} shows $\pi^0$ $A_{LL}$ preliminary
results in run 2006 at $\sqrt{s}=$62.4GeV\cite{PHENIXAokiSpin2006}.
Significant improvement can be seen in the large $x_T$ region.
Probed Bjorken-$x$ roughly scales with $x_T$ thus the results can probe
higher $x$, where we have large uncertainty\cite{AAC}.

In both cases, $\pi^0$ $A_{LL}$ is not sensitive to the sign of $\Delta G$
in the lower $x_T$ ($xT< 0.05$) due to the dominance of gluon-gluon scattering.
Measurement in higher $x_T$ where quark-gluon dominates,
or measurement of direct photon, created by quark-gluon Compton scattering,
or measurement of $\pi^{\pm}$ which have different fraction of subprocesses,
are important for the sign determination.
Other channels are also important for systematic study.

\section{Summary and Outlook}
PHENIX longitudinal spin physics program has been launched to investigate
proton spin structure. PHENIX $\pi^0$ cross section agrees well with pQCD
at both $\sqrt{s}=200$GeV and $\sqrt{s}=62.4$GeV within theoretical uncertainties
.
PHENIX $\pi^0$ $A_{LL}$ results at both $\sqrt{s}=200$GeV and
$\sqrt{s}=62.4$GeV are also shown.
They reject large $\Delta G$ senarios. Detailed comparison with
theroy is on-going at $\sqrt{s}=62.4$GeV.
Results of $\pi^0$ $A_{LL}$ at $\sqrt{s}=200$GeV in run 2006 will be available soon.
To have better sensitivity for the sign of $\Delta G$, measurement of $\pi^0$ $A_{LL}$ at
higher $x_T$, direct photon $A_{LL}$ and $\pi^{\pm}$ $A_{LL}$ are important and will be available in the future.
Other channels are also important for systematic study.
They together give us furthur information on proton spin.



\end{document}